\documentclass[acmlarge]{acmart}

\usepackage{subcaption} %

\usepackage{graphicx} %

\usepackage{algorithm}
\usepackage{algpseudocode}

\usepackage{tabularx,booktabs}

\usepackage{mathtools} %

\DeclarePairedDelimiter\ket{\lvert}{\rangle}
\DeclarePairedDelimiterX\braket[2]{\langle}{\rangle}{#1\,\delimsize\vert\,\mathopen{}#2}

\usepackage{amsmath} %

\AtBeginDocument{%
  }

\setcopyright{acmlicensed}
\copyrightyear{2024}
\acmYear{2024}
\acmDOI{XXXXXXX.XXXXXXX}

\acmJournal{TQC}
\acmVolume{37}
\acmNumber{4}
\acmArticle{111}
\acmMonth{8}

\begin{document}

\title[Incorporating Quantum Advantage in Quantum Circuit Generation]{Incorporating Quantum Advantage in Quantum Circuit Generation through Genetic Programming}

\author{Christoph Stein}
\email{christoph.stein@student.kit.edu}
\affiliation{%
  \institution{Karlsruhe Institute of Technology}
  \city{Karlsruhe}
  \country{Germany}
}

\author{Michael Färber}
\email{michael.faerber@tu-dresden.de}
\affiliation{%
  \institution{TUD Dresden University of Technology}
  \city{Dresden}
  \country{Germany}
}

\begin{abstract}
Designing efficient quantum circuits that leverage quantum advantage compared to classical computing has become increasingly critical. Genetic algorithms have shown potential in generating such circuits through artificial evolution. However, integrating quantum advantage into the fitness function of these algorithms remains unexplored. In this paper, we aim to enhance the efficiency of quantum circuit design by proposing 
two novel approaches for incorporating quantum advantage metrics into the fitness function of genetic algorithms.\footnote{The code for this paper is available at \url{https://github.com/westoun/gp4qc}.} We evaluate our approaches based on 
the Bernstein-Vazirani Problem and the Unstructured Database Search Problem as test cases. The results demonstrate that our approaches not only improve the convergence speed of the genetic algorithm but also produce circuits comparable to expert-designed solutions. Our findings suggest that automated quantum circuit design using genetic algorithms that incorporate a measure of quantum advantage is a promising approach to accelerating the development of quantum algorithms. 
\end{abstract}

\begin{CCSXML}
<ccs2012>
   <concept>
       <concept_id>10010583.10010786.10010813.10011726</concept_id>
       <concept_desc>Hardware~Quantum computation</concept_desc>
       <concept_significance>500</concept_significance>
       </concept>
   <concept>
       <concept_id>10010520.10010521.10010542.10010550</concept_id>
       <concept_desc>Computer systems organization~Quantum computing</concept_desc>
       <concept_significance>500</concept_significance>
       </concept>
   <concept>
       <concept_id>10010147.10010257.10010293.10011809.10011812</concept_id>
       <concept_desc>Computing methodologies~Genetic algorithms</concept_desc>
       <concept_significance>500</concept_significance>
       </concept>
   <concept>
       <concept_id>10002944.10011123.10010912</concept_id>
       <concept_desc>General and reference~Empirical studies</concept_desc>
       <concept_significance>300</concept_significance>
       </concept>
 </ccs2012>
\end{CCSXML}

\ccsdesc[500]{Hardware~Quantum computation}
\ccsdesc[500]{Computer systems organization~Quantum computing}
\ccsdesc[500]{Computing methodologies~Genetic algorithms}
\ccsdesc[300]{General and reference~Empirical studies}

\keywords{Quantum Computing, Genetic Programming, Genetic Algorithms, Quantum Circuit Generation, Quantum Advantage}

\received{[TO BE FILLED]}
\received[revised]{[TO BE FILLED]}
\received[accepted]{[TO BE FILLED]}

\maketitle

\section{Introduction}
\label{introduction}

In contrast to classical computing, quantum computing makes explicit 
use of phenomena that only exist in the quantum realm, such as 
superposition states and quantum entanglement. Based on these differences, 
quantum algorithms have been developed that show significant improvements 
in time complexity compared to their best performing classical 
counterparts (e.g., \cite{grover1996fast, shor1999polynomial, deutsch1985quantum}). 
Despite their potential, however, comparatively few quantum algorithms with 
a significant quantum advantage have been found over the years \cite{shor2003haven, spector1998genetic, uprety2020survey}.
One promising approach to bridge this gap is through the use
of genetic algorithms. 
These algorithms have shown to be capable of producing quantum
circuits for various tasks from entanglement production \cite{bautu2007quantum, rubinstein2001evolving}
to algorithm development \cite{ahsan2020autoqp, massey2006human, spector1998genetic}
and quantum error correction \cite{tandeitnik2022evolving}.

Within the literature on quantum circuit generation through the
use of genetic algorithms, approaches mainly differ in the way they 
represent individual quantum circuits, the quantum gates they
utilize, and how they measure the fitness of a generated 
circuit or algorithm (see Sec.~\ref{related}).
This last aspect, the fitness function, 
determines the shape of the search space the genetic algorithm
will seek to find the optimal circuit or algorithm within.
Multiple authors defined their fitness 
function to not just include the precision of the generated 
circuits on the chosen target task, but also their complexity and 
gate count
\cite{spector1998genetic, tandeitnik2022evolving, gemeinhardt2023hybrid}.

However, while previous work successfully showed the feasibility of 
learning valid quantum circuits through genetic algorithms
for a wide range of studied problems \cite{ahsan2020autoqp, bautu2007quantum}, to the best of our knowledge, no approach exists that incorporates the measurement of %
quantum advantage in the genetic algorithm itself.
Even if the generated circuits are able to produce the 
desired target states of a problem, by omitting the use of 
superposition and entanglement, they can fail to produce 
an advantage over classical non-quantum solutions in terms of 
their time or space complexity.

In this paper we investigate whether 
incorporating a measure for the quantum advantage in the
fitness function of a genetic algorithm has an impact
on the quality of the 
generated circuits and the convergence behavior of the 
genetic algorithm. 
To evaluate our approach, we chose the Bernstein-Vazirani Problem \cite{bernstein1993quantum} 
and the Unstructured Database Search Problem \cite{grover1996fast} as test studies.
Since quantum algorithms with a proven quantum advantage 
exist for each of these test problems, generated circuits 
can be compared against the state of the art in regard to 
their performance, circuit architecture, and complexity.
While most work on the use of genetic algorithms for quantum circuit generation is evaluated based on the resulting states of the produced quantum circuits, we additionally evaluate with respect to convergence behavior and speed across multiple experiment rounds to assess the robustness of our approaches. 

In short, we make the following main contributions in the paper: 
\begin{itemize}
    \item 
    We propose a novel fitness function for genetic algorithms that incorporates an explicit measure of the degree to which quantum advantage is achieved. 
    \item We propose a second fitness function that incorporates quantum advantage indirectly based on the differences between quantum and classical computation. 
    \item We evaluate our approaches based on two established 
    test studies, the Bernstein-Vazirani Problem \cite{bernstein1993quantum} and the Unstructured Database Search Problem \cite{grover1996fast}, and report the best performing circuits generated by each fitness function as well as the convergence behavior of the genetic algorithm.
\end{itemize}

The rest of this paper is structured as follows:
Sec. \ref{related} contains an overview of related work in the field of 
quantum circuit generation through genetic algorithms.
Sec. \ref{method} elaborates on the suggested approach and introduces 
the proposed fitness functions in detail.
The results of experiments run on the Bernstein-Vazirani Problem
as well as the Unstructured Database Search Problem are presented 
in Sec. \ref{results}. 
Sec. \ref{discussion} distills the implications of the afore presented 
results and Sec. \ref{conclusion} concludes and points to future work. 

\section{Related Work}
\label{related}

Multiple techniques have been developed
with the goal of automatically generating quantum
circuits and quantum algorithms. 
While recent work often makes use of reinforcement learning to 
select which gates to add to a circuit \cite{kuo2021quantum, mckiernan2019automated}, 
a vast body of work exists that uses genetic algorithms 
to generate quantum circuits for various tasks and performance goals. %
Since this paper builds on recent work in the context of 
genetic algorithms, we focus on this aspect in the following paragraphs. %

\begin{table}[tb]
\centering 
\caption{Related approaches of genetic algorithms for quantum circuit design, sorted by publication year.}
\label{tab:fitness-functions}
\begin{tabular}{>{\raggedright\arraybackslash}p{9em} >{\raggedright\arraybackslash}p{22em} >{\raggedright\arraybackslash}p{11em}} 
  \toprule
  \textbf{Authors} & \textbf{Fitness Function} & \textbf{Use of Parameterized Gates} \\ 
  \midrule

  Gemeinhardt et al. \cite{gemeinhardt2023hybrid} & Multi-objective including accuracy, gate count, and circuit depth. & Nelder-Mead optimization. \\ 

  Tandeitnik et al. \cite{tandeitnik2022evolving} & Custom quality score combined with circuit depth. & None. \\

  Franken et al. \cite{franken2022quantum} & Deviation between realized and target Hamiltonian. & Mutation operator. \\

  Ahsan et al. \cite{ahsan2020autoqp} & Absolute deviation between target and desired state. & None. \\ 

  Bautu et al. \cite{bautu2007quantum} & Absolute deviation between target and desired state. & None. \\

  Massey et al. \cite{massey2006human} & Magnitude and phase combination, with an efficiency term. & Mathematical operators. \\

  Massey et al. \cite{massey2004evolving} & Adapted from Spector et al. with threshold adjustments. & As per Spector et al. \\

  Rubinstein et al. \cite{rubinstein2001evolving} & Normalized deviation between target and desired state. & Randomized rotation parameters ($[-\pi, \pi]$). \\ 

  Williams et al. \cite{williams1998automated} & Absolute deviation between desired and realized unitary. & None. \\

  Spector et al. \cite{spector1998genetic} & Multi-step fitness involving hits, deviation, and circuit length. & Mathematical operators. \\

 \bottomrule
\end{tabular}
\end{table}

\subsection{Fitness Functions}

The fitness function serves as the basis of selection 
within a genetic algorithm. 
Its choice determines which 
individuals are more likely to survive and, hence, which traits
are more likely to spread throughout the population. 
Since most quantum circuits are developed to fulfill a specific
computational need, the difference between 
the actual and the desired final state of the quantum circuit forms 
the basis of most fitness functions used in the literature. 
Table \ref{tab:fitness-functions} provides an overview of the fitness functions and use of parameterized gates of related genetic algorithms for quantum circuit design.

Bautu et al. \cite{bautu2007quantum} and 
Ahsan et al. \cite{ahsan2020autoqp} use the 
absolute deviation between the actual and the target state of a 
quantum circuit as its fitness value, while 
\cite{williams1998automated} utilizes the absolute deviation 
as the starting point for a custom ranking approach. 
Gemeinhardt~\cite{gemeinhardt2023hybrid} uses a combination of accuracy, 
gate count, parameter count, and other measures in connection
with the multi-objective optimization algorithm NSGA-III 
\cite{deb2013evolutionary} to account for the trade-offs between
circuit performance and circuit complexity.
\cite{tandeitnik2022evolving} also incorporates the trade-off 
between performance and complexity, although they do so
through a weighted combination
of a quality measure and circuit depth. 
While the above mentioned approaches apply both a measure for the 
performance and complexity of the generated circuits
in their fitness functions,
to the best of our knowledge no approach 
exists that incorporates a measure of quantum advantage,
the driving force behind quantum algorithm development.

Acknowledging the probabilistic nature of quantum computing,  
Spector et al. \cite{spector1998genetic} developed a custom 
multi-component fitness function, further refined by 
\cite{massey2004evolving}.
The first component, hits, rewards any additional test case 
where the probability of producing the correct solution is 
above a threshold of 0.52 (0.5 + 0.02 to account for rounding 
errors). 
The second component, correctness, adds an error term for each 
incorrect test case based on the difference between the desired
result and the produced distribution. 
Finally, if a circuit passes all test cases, an efficiency term
is applied based on the amount of gates used by a circuit. 
In total, this fitness function first drives the population 
towards optimizing the amounts of correct test cases, nudging 
it to reduce circuit complexity from that point on.
The fitness functions introduced in this work 
can be considered as extensions of the 
fitness function of Spector et al. \cite{spector1998genetic}, 
further incorporating a 
measure of quantum advantage into the optimization procedure.

\subsection{Gate Types and Parameterized Gates}

Regarding the types of gates used, related
work can be dichotomized based on its use or avoidance of 
parameterized gates.
While all gates have at least one parameter describing which 
qubit(s) they apply to, some gates, like the phase gate, 
take additional parameters influencing the scope of the 
manipulation they perform.
These gates usually correspond to controlled rotations about
one or more axes, with their parameters falling in the range 
of $[-\pi, \pi]$.

Since these parameterized gates allow for a fine control of 
the states of one or more qubits, they form the basis of many 
advanced quantum algorithms like the variational quantum 
eigensolver \cite{peruzzo2014variational} or the quantum fourier transform \cite{coppersmith2002approximate}. 
For their use in the context of genetic programming, 
suitable parameters for those gates have to be 
learned.
While tree-based approaches like \cite{spector1998genetic} and 
\cite{massey2004evolving} incorporate parameter 
optimization through the use of mathematical operators in their
representation scheme, other approaches like 
\cite{gemeinhardt2023hybrid} incorporate parameter optimization 
as an additional step in their genetic algorithm. 
\cite{franken2022quantum} and \cite{giovagnoli2023qneat} adjust 
rotation parameters through the use of an additional mutation 
operator within the genetic algorithm itself.

The approach presented in this work will follow the work of 
Gemeinhardt et al. \cite{gemeinhardt2023hybrid} and include 
parameter optimization as an additional step in the genetic 
algorithm.
While this does require additional computational resources
during the execution, it greatly enhances the 
effectiveness of the algorithm on complex problems. %

\section{Method}
\label{method}

The goal of this work is to develop a genetic 
algorithm for quantum circuit generation that 
incorporates in its fitness function whether or not a quantum 
advantage is achieved.
Two proposed fitness functions, each approaching 
this objective from a different angle, are introduced and defined in 
Sec. \ref{method:direct} and Sec. \ref{method:indirect}
respectively.

\subsection{Direct Incorporation of Quantum Advantage}
\label{method:direct}

Most established quantum algorithms define their quantum 
advantage based on the amount of black box oracle calls
they require to solve a computational problem.
This number is then contrasted 
to the amount of oracle calls the best performing classical (non-quantum)
algorithm would need \cite{deutsch1985quantum, grover1996fast, bernstein1993quantum}.
In the case of the Unstructured Database Search Problem, for 
example, the best performing classical algorithm would require on average 
$N/2$ oracle calls to find a specific entry within an unstructured
list.
Grover's Algorithm \cite{grover1996fast}, in contrast, has been 
shown to require only $O(\sqrt{N})$ oracle calls.
In this section, we introduce a fitness function that incorporates
this respective speedup, the quantum advantage a generated circuit achieves in contrast
to its best performing classical counterpart. 

In the work of Spector et al. \cite{spector1998genetic}, 
the fitness function comprises of 
multiple factors that each target different aspects of the circuits
to be created. 
At first, each circuit is shaped to solve as many test cases 
as possible.
Then, once the algorithm finds a circuit that passes all test cases, 
the fitness function further optimizes by preferring circuits 
that make use of fewer gates.

While the amount of gates used serves as an approximate measure 
of the simplicity of a circuit, it is an insufficient indicator of 
the degree of quantum advantage achieved. 
In contrast to the amount of oracle calls employed, 
the amount of gates used by a circuit does not indicate
the scaling behavior of the underlying algorithm.
Furthermore, the amount of gates used for implementing an algorithm
heavily depends on the gate set used, a concern driven by 
hardware and design considerations,
irreverent of the underlying algorithm itself.

In this work, in addition to incorporating the amount of a gates 
as a factor indicating circuit simplicity, we incorporate a custom 
measure
for the degree of quantum advantage achieved by a circuit in its
fitness value computation. 

\begin{algorithm}
\caption{The fitness function of Spector et al. \cite{spector1998genetic}, adjusted to directly incorporate a measure of quantum advantage.} 

\begin{algorithmic}[1]
    \State $probabilities \gets evaluate(circuit)$

    \State $hits \gets \#TestCases$
    \State $error \gets 0$

    \For{i = 0; i < \#TestCases; i++}

        \State $j \gets getTargetIndex(targetProbabilities[i])$
    
        \If{probabilities[i][j] >= 0.52}
        
            \State $hits \gets hits - 1$
            
        \Else
            
            \State $\delta \gets jensenShannonDistance(probabilities[i], targetProbabilities[i])$
            
            \State $error \gets error + \delta$
        \EndIf
    \EndFor

    \If{hits > 0}
        
        \State $fitness \gets hits + \frac{error}{max(hits, 1)}$
        
    \Else 

        \State $fitness \gets \frac{countOracleGates(circuit)}{classicalOracleCalls} + 
        \frac{countGates(circuit)}{100000} $

    \EndIf

    \State \Return fitness
\end{algorithmic}
\label{alg:directqa_fitness}
\end{algorithm}

Algorithm \ref{alg:directqa_fitness} shows the first novel fitness
function proposed in this paper. 
It iterates over all possible test cases (input values or oracle 
implementations) and compares the probability distributions generated 
by the current circuit with the desired.
If the desired target state for each test case is measured with 
a probability of at least 0.52 (based on \cite{spector1998genetic}), 
the hits counter that measures the amount of incorrect test cases is 
reduced by one.
If the produced state does not meet this threshold for a test case, 
the Jensen Shannon Distance between the actual and desired 
probability distribution is computed and added to the existing error 
term. 
After all test cases have been evaluated, the fitness score is computed 
based on the amount of incorrect test cases measured by the hits counter.
If at least one test case has not been passed, the fitness score
is defined as the sum of the amount of remaining test cases and the 
average error term of these cases. 
If all test cases have been passed, the fitness score is computed based
on the amount of gates used in the circuit 
and the relation of quantum oracle calls to classical oracle calls.

Using this fitness function, the genetic algorithm is driven to 
first improve the 
amount of correct test cases (line 13-15 in Alg. \ref{alg:directqa_fitness}). 
Then, once all test cases have been cleared, an additional 
fitness
term is applied that serves as a punishment factor for the amount
of quantum oracle calls used (line 16 in Alg. \ref{alg:directqa_fitness}).
Since many quantum algorithms define the degree of quantum advantage 
achieved based on the lower amount of oracle calls needed, we divide
the amount of oracle calls used in a quantum circuit by the amount of 
oracle calls the best performing classical (non-quantum) algorithm would require. 

The fewer oracle calls a quantum circuit uses in relation to the 
best performing classical algorithm, the lower its comparative 
fitness value
and the higher its likelihood of surviving selection.
Thereby, circuits that solve all test cases but use  
fewer oracle calls are more likely to move on to the next generation.
In the end, these circuits show a higher degree of quantum 
advantage.
The remaining term of Spector's original fitness function, 
efficiency, is left as it was originally defined.

\subsection{Indirect Incorporation of Quantum Advantage through Constraints}
\label{method:indirect}

As shown in \cite{bernstein1993quantum}, any computation 
a quantum 
computer can do can also be performed by a classical (non-quantum) computer. 
The difference between quantum and non-quantum computers 
lies not in \textit{what} they are able do, but in \textit{how 
efficiently} they are able to do it.
Quantum computers are able to exhibit quantum 
advantages because in a quantum computer superposition and entanglement
happen at the physical level. 
Thus, the fitness function introduced in this section punishes all 
circuits that do not make use of entanglement and superposition,
since such circuits cannot exhibit significant quantum advantage.

From a quantum circuit perspective, multiple gates can be used 
to create a state of \textit{superposition}.
Each of these gates transforms a qubit from one of its base states 
to a gate-specific superposition state, thereby fulfilling the 
proposed superposition condition.
While the most obvious candidate, the Hadamard gate, creates an
equal superposition state between both basis states, rotational 
gates targeting the x- or y-axis can also be used to create 
superposition states.
Furthermore, same holds for the controlled versions of each 
of those gates.

To achieve a state of \textit{entanglement} in which the states of individual 
qubits cannot be described independently of one another, the 
state of one qubit has to depend on the state of another. 
In the quantum circuit representation, this resembles the 
idea of controlled gates, where a specific operation is only
applied to a qubit if the corresponding control qubit is 
in a specific state. 
In the case of the CNOT gate, for example, the Pauli-X gate is 
applied to the target qubit iff the control qubit is in state 
$\ket*{1}$.
The degree of entanglement between the qubits of a system can 
thereby be increased by applying controlled gates to previously
un-entangled qubits.

\begin{algorithm}
\caption{The fitness function of Spector et al. \cite{spector1998genetic}, adjusted to incorporate the potential for quantum advantage through the aforementioned constraints.} 

\begin{algorithmic}[1]
    \State $probabilities \gets evaluate(circuit)$

    \State $hits \gets \#TestCases$
    \State $error \gets 0$

    \For{i = 0; i < \#TestCases; i++}
        
        \State $j \gets getTargetIndex(targetProbabilities[i])$
    
        \If{probabilities[i][j] >= 0.52}
        
            \State $hits \gets hits - 1$
            
        \Else
        
            \State $\delta \gets jensenShannonDistance(probabilities[i], targetProbabilities[i])$
            
            \State $error \gets error + \delta$
        \EndIf
    \EndFor

    \State $fitness \gets 0$

    \If{not containsSuperpositionGates(circuits)}
        \State $fitness \gets fitness + (\#TestCases + 1)$
    \EndIf
    
    \If{not containsEntanglementGates(circuits)}
        \State $fitness \gets fitness + (\#TestCases + 1)$
    \EndIf

    \If{hits > 0}
        
        \State $fitness \gets fitness + hits + \frac{error}{max(hits, 1)}$
        
    \Else 

        \State $fitness \gets fitness + \frac{countGates(circuit)}{100000} $

    \EndIf

    \State \Return fitness
\end{algorithmic}
\label{alg:indirectqa_fitness}
\end{algorithm}

Similar to our approach proposed in 
Sec. \ref{method:direct}, we incorporate in this subsection the aforementioned 
quantum-specific differences in the genetic algorithm as aspects of the fitness function. 
In contrast to Sec. \ref{method:direct}, however, these 
factors are incorporated as constraints \textit{prior} to the 
optimization of fitness test cases (see lines 14-19 in Alg. \ref{alg:indirectqa_fitness}).
By opting the genetic algorithm to first create circuits with 
the potential of a quantum advantage, the algorithm will deprioritize 
any 
circuit that might solve the problem but does not include any
gates incorporating the potential for quantum advantage.

Considering that the maximum value of the baseline fitness function defined 
in 
\cite{spector1998genetic} is equal to the amount of test
cases used, the punishment terms for violating the 
aforementioned constraints shall take a value 
slightly above this maximum, in this case $\#TestCases + 1$ 
(lines 14-19 in Alg. \ref{alg:indirectqa_fitness}).
That way, the genetic algorithm is incentivized to optimize 
constraint satisfaction first, before improving performance on the 
specific problem set. 
The absolute value of the punishment term does not matter, as long 
as it is large enough to avoid trade-offs against other 
fitness components down the line.

\section{Evaluation} %
\label{results}

To evaluate the approaches proposed,
we chose the Bernstein-Vazirani Problem \cite{bernstein1993quantum}
and the Unstructured Database Search Problem \cite{grover1996fast}
as test studies.
For each of the test studies, experiments were run with 
the fitness function of Spector et al. \cite{spector1998genetic} (\textit{BaselineFitness}), 
our fitness function proposed in Sec. \ref{method:direct} (\textit{DirectQAFitness}),
and our fitness function proposed in Sec. \ref{method:indirect} (\textit{IndirectQAFitness}).
Since genetic algorithms make strong use of random choices 
both during population initialization as well as during 
each generation, each experiment configuration was run 12 times.

During each experiment run, the population size was set to 1,000, the
probability for crossover was set to 0.4, the probability of 
randomly swapping individual gates was set to 0.03, and elitism 
was performed with 10\% of the population.
These parameters were chosen based on experience and systematic 
trial and error ahead of the actual experiment runs.
On the Unstructured Database Search Problem, the chromosome length
was set to 30 and each algorithm was run for 800 generations.
On the Bernstein-Vazirani Problem, the chromosome length
was set to 15 and each algorithm was run for 500 generations.
Different parameters have been chosen for each of the test studies 
to account for the difference in complexity between both problems.

\subsection{Results on the Bernstein-Vazirani Problem}
\label{results:bernstein}

\begin{figure}[tb]
    \centering 
    \subfloat[Subfigure 1 list of figures text][Average Mean Fitness]{
        \includegraphics[width=0.4\textwidth]{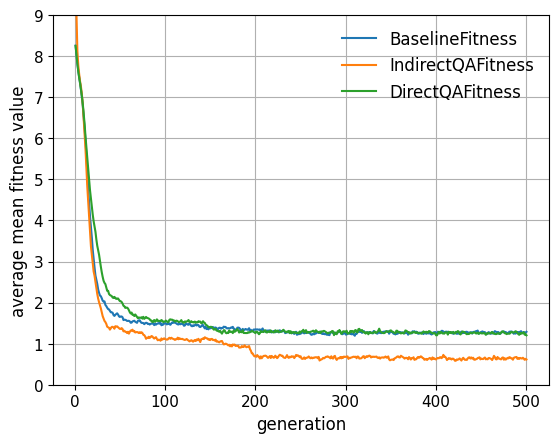}
    }
    \qquad
    \subfloat[Subfigure 2 list of figures text][Average Minimum Fitness]{
        \includegraphics[width=0.4\textwidth]{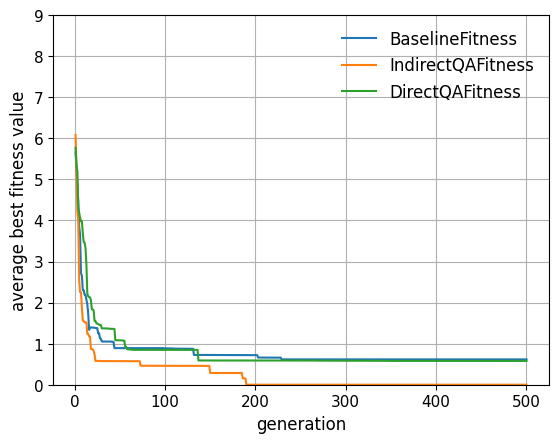}
    }
    \caption{Averaged fitness values on the Bernstein-Vazirani Problem.}
    \label{fig:bernstein_fitness_function_averaged}
\end{figure}

Fig. \ref{fig:bernstein_fitness_function_averaged} shows the development of 
the mean and minimum fitness values for each generation of the genetic algorithm, averaged across multiple 
experiment runs, and for each of the fitness functions used. 
\textit{BaselineFitness} refers to the fitness function proposed by Spector
et al. in \cite{spector1998genetic}.

While \textit{BaselineFitness} and \textit{DirectQAFitness} show 
similar convergence behavior, both in regard to their averaged mean as 
well as their averaged minimum fitness values, the fitness values 
for \textit{IndirectQAFitness} converge faster and reach lower overall
fitness values.
The higher fitness values of \textit{IndirectQAFitness} during early
generations can be explained through the punishment terms applied in
the fitness value computation of this fitness function (s. Sec. \ref{method:indirect}).
Once the superposition and entanglement constraints have been 
satisfied, \textit{IndirectQAFitness} outperforms the other two 
fitness functions in terms of its average fitness values.

\begin{figure}[tb]
    \centering 
    \subfloat[Subfigure 1 list of figures text][Baseline Fitness]{
        \includegraphics[width=0.45\textwidth]{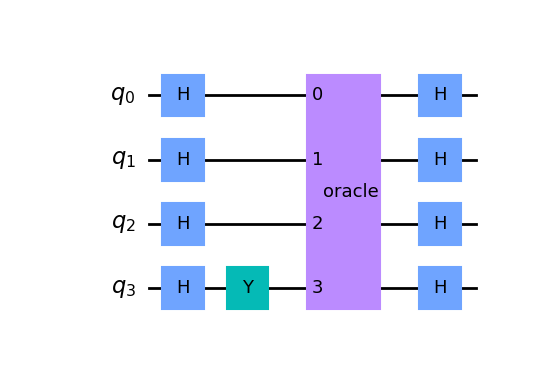}
    }
    \qquad
    \subfloat[Subfigure 1 list of figures text][Direct Quantum Advantage Fitness]{
        \includegraphics[width=0.45\textwidth]{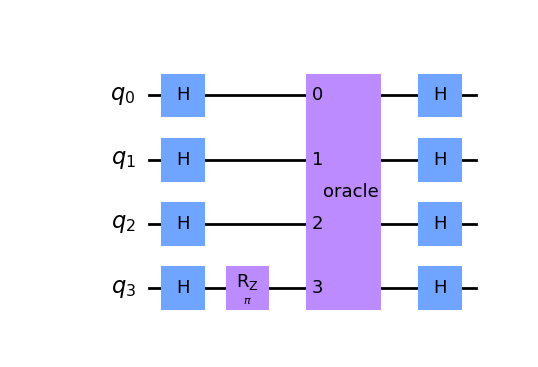}
    }
    \qquad
    \subfloat[Subfigure 1 list of figures text][Indirect Quantum Advantage Fitness]{
        \includegraphics[width=0.45\textwidth]{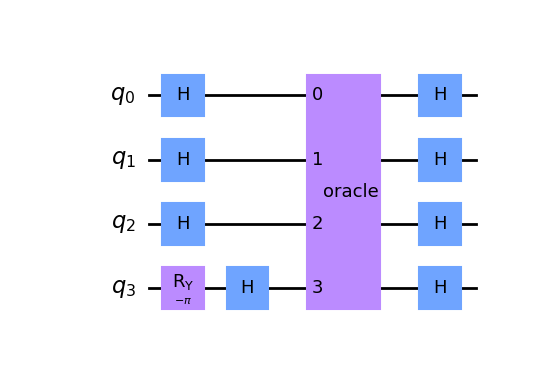}
    }
    \caption{Best performing circuits of each fitness function for the Bernstein-Vazirani Problem.}
    \label{fig:bernstein_circuits}
\end{figure}

Fig. \ref{fig:bernstein_circuits} lists the best performing circuits 
generated by each of the experiment configurations.
In this context, a circuit is considered better the lower 
its corresponding fitness value is. 
While all of the generated circuits follow the structure of 
the state of the art 
solution \cite{bernstein1993quantum}, individual circuits 
vary in the way they manipulate the ancillary qubit q3. 
Aside from a global phase factor, each of these circuits
is equivalent to the state of the art circuit of \cite{bernstein1993quantum}, 
which was created manually. 
Each of these circuits returns the desired result state with a 
probability of 1.

\subsection{Results on the Unstructured Database Search Problem}
\label{results:grover}

\begin{figure}[tb]
    \centering 
    \subfloat[Subfigure 1 list of figures text][Average Mean Fitness]{
        \includegraphics[width=0.4\textwidth]{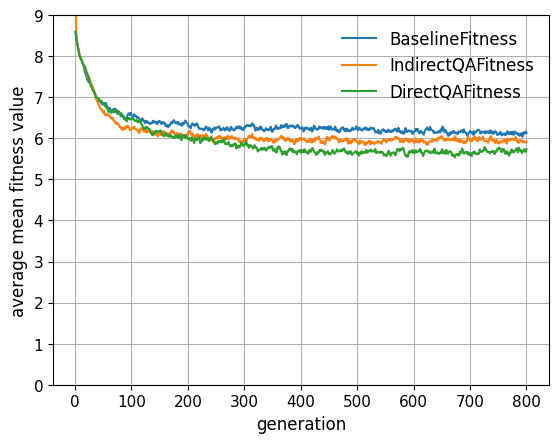}
    }
    \qquad
    \subfloat[Subfigure 2 list of figures text][Average Minimum Fitness]{
        \includegraphics[width=0.4\textwidth]{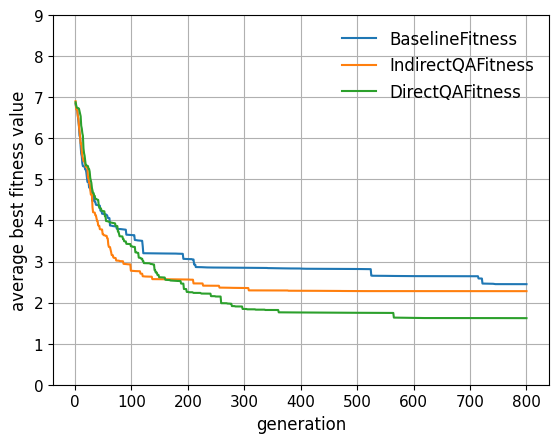}
    }
    \caption{Averaged fitness values on the Unstructured Database Search Problem, grouped by fitness function.}
    \label{fig:grover_fitness_function_averaged}
\end{figure}

Fig. 
\ref{fig:grover_fitness_function_averaged} shows a slight 
out-performance of the \textit{IndirectQAFitness} and \textit{DirectQAFitness} in comparison to the \textit{BaselineFitness} function in regards to both their averaged mean values as well as their averaged minimum fitness values.
While the \textit{IndirectQAFitness} has a higher averaged mean 
fitness value during early generations, due to the punishment terms
applied in the fitness value computation, this fitness function
converges faster, although to a less beneficial value compared 
to the \textit{DirectQAFitness} function.

In general, the averaged minimum fitness values as 
well as the minimum fitness values of each experiment run  
are monotonically decreasing, due to the 
use of elitism in the genetic algorithm.
In our implementation, elitism is realized by copying the 
best performing circuits of each generation into the next 
generation unaltered.
As a consequence, the minimum fitness value of a population can only
stay stable or decrease as generations progress.

\begin{figure}[tb]
\makebox[\textwidth][c]{
    \includegraphics[width=20cm, keepaspectratio, height=\textheight]{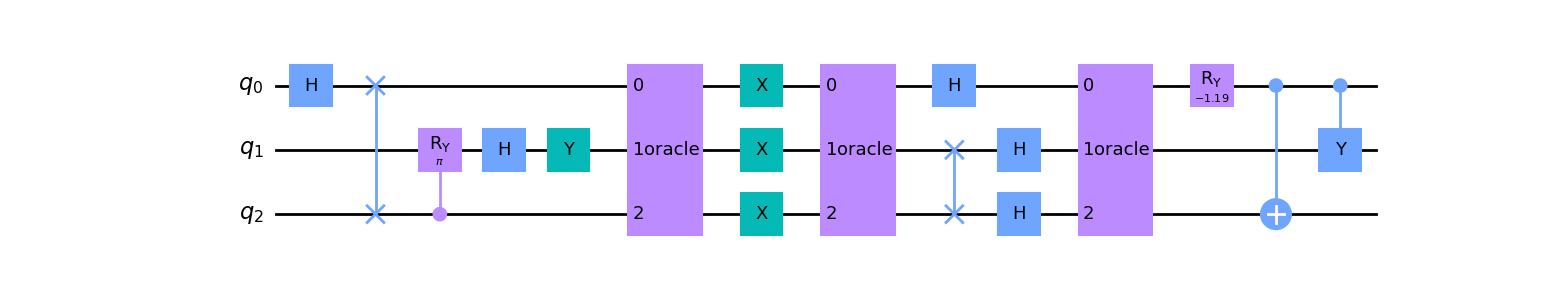}
}
    \caption{Best performing circuit for \textit{BaselineFitness} on the Unstructured Database Search Problem.}
    \label{fig:grover_circuit_baseline_no}
\end{figure}

\begin{figure}[tb]
\makebox[\textwidth][c]{
    \includegraphics[width=18cm, keepaspectratio, height=\textheight]{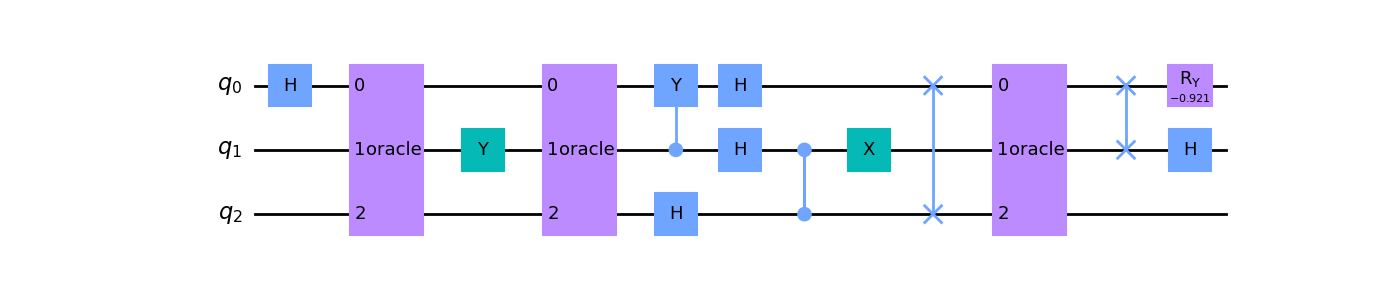}
}
    \caption{Best performing circuit for \textit{DirectQAFitness} on the Unstructured Database Search Problem.}
    \label{fig:grover_circuit_directqa_no}
\end{figure}

\begin{figure}[tb]
\makebox[\textwidth][c]{
    \includegraphics[width=15cm, keepaspectratio, height=\textheight]{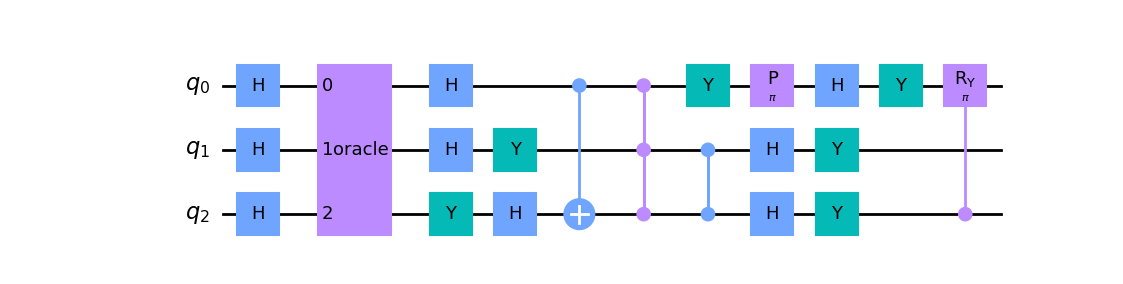}
}
    \caption{Best performing circuit for \textit{IndirectQAFitness} on the Unstructured Database Search Problem.}
    \label{fig:grover_circuit_indirectqa_no}
\end{figure}

Figures \ref{fig:grover_circuit_baseline_no} to \ref{fig:grover_circuit_indirectqa_no}
show the best performing quantum
circuits across all runs of each of the experiment configurations. 
Each of these circuits achieved a fitness value close 
to 0, indicating that it passed all possible test cases of the Unstructured 
Database Search Problem on 3 qubits.

In the following, we 
provide a case study by discussing specific  
parallels and differences between the generated best performing circuits %
and the state of the art solution from \cite{grover1996fast}.

In addition to employing the same amount of oracle calls, 
the circuits in Fig. \ref{fig:grover_circuit_baseline_no} and Fig. \ref{fig:grover_circuit_directqa_no} follow a shared 
structure that is notably structurally different from the state of the art circuit.
Instead of creating an equal superposition at the beginning of the 
circuit, as the state of the art solution does, these circuits 
start by placing the system in a partial superposition state.
The circuit illustrated in Fig. \ref{fig:grover_circuit_directqa_no},
for example, begins by putting only the first qubit in
superposition. 
Since the state of the system after the first Hadamard gate is 
$\ket*{\psi} = \frac{\ket*{000} + \ket*{100}}{\sqrt{2}}$, the first
oracle call only produces an effect, if the desired target state of the oracle
follows the schema $\ket*{?00}$, where \textit{$?$} serves as a placeholder
for either $\ket*{0}$ or $\ket*{1}$.
If the second or the third qubit of the target state is in a state of 
$\ket*{1}$, the first oracle
call does not have any effect.
Similarity, the second oracle calls only has an effect, if the desired
target state follows the schema $\ket*{?10}$, since the Y gate acts
on the second qubit by flipping it from $\ket*{0}$ to $\ket*{1}$ while 
adding a global phase value of $-i$.
Following the second oracle call, a superposition over all possible states 
is created through the use of multiple Hadamard gates. 

In contrast to circuits in Fig. \ref{fig:grover_circuit_baseline_no} and 
Fig. \ref{fig:grover_circuit_directqa_no}, the circuit of Fig. 
\ref{fig:grover_circuit_indirectqa_no}
makes use of only a single oracle call.
It starts by creating an equal superposition of all 
possible states using Hadamard gates, similar to the
state of the art \cite{grover1996fast}.
Of special interest is the unitary matrix of the gates following 
the oracle call in Fig. \ref{fig:grover_circuit_indirectqa_no}:

\begin{center}
    \begin{math}
    Unitary_{Circuit Fig. \ref{fig:grover_circuit_indirectqa_no}} = \begin{pmatrix}
        -0.75 & 0.25 & 0.25 & 0.25 & 
        0.25 & 0.25 & 0.25 & 0.25\\
        
        -0.25 & 0.75 & -0.25 & -0.25 & 
        -0.25 & -0.25 & -0.25 & -0.25\\
        
        0.25 & 0.25 & -0.75 & 0.25 & 
        0.25 & 0.25 & 0.25 & 0.25\\
        
        -0.25 & -0.25 & -0.25 & 0.75 & 
        -0.25 & -0.25 & -0.25 & -0.25\\
        
        0.25 & 0.25 & 0.25 & 0.25 & 
        -0.75 & 0.25 & 0.25 & 0.25\\
        
        0.25 & 0.25 & 0.25 & 0.25 & 
        0.25 & -0.75 & 0.25 & 0.25\\
        
        0.25 & 0.25 & 0.25 & 0.25 & 
        0.25 & 0.25 & -0.75 & 0.25\\
        
        0.25 & 0.25 & 0.25 & 0.25 & 
        0.25 & 0.25 & 0.25 & -0.75
    \end{pmatrix}
\end{math}
\end{center}

Aside from the sign of row 2 and 4, this matrix is exactly equal to 
the diffusion operator described by Lov Grover in 
\cite{grover1996fast}.
Since no operations take place after the gates composing this 
operator have been applied, the signs of its rows do not have an
effect on the observed probabilities.
Without any knowledge of or limitations in regard to the gate set 
used by the state of the art solution, the genetic algorithm was able
to construct a complex operator similar to the state of the art in 
both functionality and form.

\section{Main Findings}
\label{discussion}

In regards to the main contributions of this paper, we found that:

\begin{itemize}
    \item The results from both test studies demonstrate that the proposed fitness functions result in faster and more effective convergence on average, compared to the baseline fitness function from \cite{spector1998genetic}. This suggests that incorporating a measure of quantum advantage enhances the genetic programming approach, enabling the generation of more effective quantum circuits within fewer generations.
    
    \item Each experimental configuration successfully produced circuits for the Bernstein-Vazirani Problem that are similar in structure and performance to state-of-the-art solutions. The differences were minimal, with variations typically involving only a single gate, while the overall structure remained consistent with established methods.
    
    \item In the more complex Unstructured Database Search Problem, the best-performing circuits exhibited greater variability in both structure and gate selection. Notably, two of the top circuits shared similarities in their entanglement strategies and oracle call usage, offering an effective alternative to the state-of-the-art approach by utilizing a step-wise entanglement procedure.
    
    \item Among the circuits generated for the Unstructured Database Search Problem, the one illustrated in Fig.~\ref{fig:grover_circuit_indirectqa_no} closely mirrors the diffusion operator described by \cite{grover1996fast}. This circuit effectively redistributes probability amplitudes towards the marked state, functioning similarly to Grover's diffusion operator. Remarkably, the genetic algorithm independently generated a circuit that aligns closely with the state-of-the-art solution in both performance and interpretability, despite lacking prior knowledge of the existing solution.
\end{itemize}

\section{Conclusion}
\label{conclusion}

We proposed two novel fitness functions, each taking a different route
to incorporate a measure of quantum advantage in the genetic 
algorithm.
When evaluated on the Bernstein-Vazirani Problem and the Unstructured
Database Problem, the suggested fitness functions outperformed the 
baseline fitness function of Spector et al. \cite{spector1998genetic}
in terms of convergence speed and the quality of the produced circuits.
While the experiments run in the context of this paper have been 
able to closely 
reproduce the state of the art solution on the Bernstein-Vazirani 
Problem, they yielded interesting new circuit architectures on the 
Unstructured Database Problem.
In one of the experiments, a circuit was learned that closely 
resembles the diffusion
operator suggested by Lov Grover in \cite{grover1996fast}.
These findings affirm the overall potential of genetic algorithms for 
the discovery and generation of meaningful new quantum algorithms and 
circuit architectures.

In future work, we plan to incorporate a measure of well-formedness for the probability amplitude distributions generated by each quantum circuit. Currently, all oracle cases are treated independently. However, since quantum oracles are designed as black boxes, quantum algorithms should exhibit consistent behavior across different oracle cases and implementations. If the probability amplitude distributions produced by a circuit vary significantly across different oracle implementations, it suggests that the circuit may be overfitting to specific test cases rather than addressing the underlying abstract problem. Exploring how to integrate a well-formedness measure into the fitness function of the genetic algorithm, and assessing whether this adjustment impacts the quality and convergence of the algorithm, represents a promising direction for future research.

\bibliographystyle{ACM-Reference-Format}
\bibliography{PAPER.bib}

\appendix

\end{document}